\documentstyle[11pt,moriond,epsfig,rotate]{article}

\def\be{\begin{equation}}
\def\ee{\end{equation}}
\def\bea{\begin{eqnarray}}
\def\eea{\end{eqnarray}}

\begin{document}
\vspace*{4cm}
\title{RARE CHARM DECAYS}

\author{EUGENE GOLOWICH}

\address{Department of Physics, University of Massachusetts\\
Amherst MA 01003, USA}

\maketitle\abstracts{
This paper is a written version of a talk on rare (FCNC) $D$ 
meson decays as presented at the Electroweak Moriond 2002.  
The presentation proceeds in two parts.  We first consider 
Standard Model predictions, taking into account both 
short-distance and long distance effects.  Then several 
New Physics options ({\it e.g.} supersymmatry, strong dynamics, 
extra large dimenions, {\it etc}) are considered.}  

\section{Introductory Remarks}
I dedicate this paper (as I did my talk) to 
Jean Tran Thanh Van and his wife for their valuable contributions 
over many years to our discipline. 

My talk summarized a recent paper~\cite{bghp2} written 
in collaboration with Gustavo Burdman, JoAnne Hewett and 
Sandip Pakvasa on the topic of rare flavor-changing neutral current 
(FCNC) decays of $D$ mesons.  Transitions of this class include: 
(i) $D\to V \gamma \ (V = \rho, \omega, \phi, \dots )$,
\footnote{The single-photon decays were considered previously 
by the authors in a separate paper.~\cite{bghp1}} 
(ii) $D\to X \ell^+ \ell^- \ (X = \pi, K, \eta,
\rho, \omega, \phi, \dots )$, 
(iii) $D\to X \nu_\ell {\bar \nu}_\ell \ (X = \pi, K, \eta,
\dots)$, (iv) $D\to \gamma \gamma$, 
(v) $D\to \ell^+ \ell^-$.

Perhaps the most noteworthy aspect of the above list is that, 
at the time of this writing, exactly {\it zero} FCNC events 
have been detected.~\cite{pdg}  We are optimistic that this situation 
will be rectified in the fairly near future.  The database for 
charm-related physics continues to expand, and experiments 
at B-factories, hadron collidiers and tau/charm facilities 
will soon achieve truly interesting sensitivities.  

As such, one can reasonably imagine a future conference talk 
which announces the detection of a $D$ meson FCNC transition. 
How to interpret such a signal?  There is a commonly adopted 
procedure which should be followed.  First one determines whether the 
Standard Model (SM) can explain the observed events.  If not, one 
proceeds to consider a menu of New Physics models.  

\section{Standard Model Analysis}

\begin{figure}
\vskip .1cm
\hskip 3.5cm
\rotate[l]{\psfig{figure=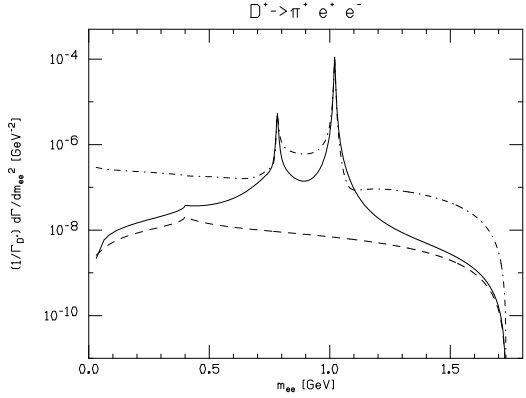,height=2.7in}}
\caption{Dilepton mass distribution for $D^+\to 
\pi^+ e^+ e^-$.  The dashed (solid) line is the short-distance 
(total) SM contribution.  The dot-dashed line is the 
R-parity violating contribution.\hfill 
\label{fig:pill}}
\end{figure}

In principle, the task of producing SM predictions for 
FCNC $D$ meson decays is straightforward.  There are 
two components to the analysis, short-distance (SD) and 
long-distance (LD), which must be separately calculated.  
We consider each in turn.
\subsection{The short-distance component}\label{subsec:sd} 
Short-distance amplitudes are concerned with 
the QCD degrees of freedom (quarks, gluons) and any relevant 
additional fields (leptons, photons).  Thus, the short distance 
part of the $D \to X_u \ell^+ \ell^-$ amplitude involves the 
quark process $c \to u \ell^+ \ell^-$.  It is usually 
most natural to employ an effective description in which the weak 
hamiltonian is expressed in terms of local multiquark operators 
and Wilson coefficients.~\cite{bbl} For example, the effective 
hamiltonian for $c\to u\ell^+\ell^-$ with renormalization scale 
$\mu$ in the range $m_b \ge \mu \ge m_c$ is~\footnote{Quantities 
with primes have had the explicit $b$-quark contributions 
integrated out} 
\begin{equation}
{\cal H}_{\rm eff}^{c\to u\ell^+\ell^-} = - {4 G_F \over \sqrt{2}}
\left[ \sum_{i=1}^2~\left(\sum_{q=d,s} ~C_i^{(q)}(\mu) 
{\cal O}_i^{(q)}(\mu) \right) + 
\sum_{i=3}^{10} ~C_i^{'}(\mu) {\cal O}_i^{'}(\mu) \right] \ \ .
\label{weakham}
\end{equation}
In the above, ${\cal O}_{1,2}^{(q)}$ are four-quark current-current 
operators, ${\cal O}^{'}_{3-6}$ are the QCD penguin operators, 
${\cal O}_7$ (${\cal O}_8$) is the electromagnetic 
(chromomagnetic) dipole operator and ${\cal O}_{9,10}$ 
explicitly couple quark and lepton currents.  For example, we 
have 
\be
O_7^{'} = \frac{e}{16\pi^2}m_c(\bar{u}_L\sigma_{\mu\nu}c_R)F^{\mu\nu}\ ,
\qquad O_9^{'} = \frac{e^2}{16\pi^2} (\bar{u}_L\gamma_\mu c_L)
( \bar{\ell}\gamma^\mu \ell) \ \ .
\label{examples}
\ee
The famous Inami-Lim functions~\cite{il} contribute to the Wilson 
coefficients $C_{7-10}$ at scale $\mu = M_{\rm W}$.

Figure~\ref{fig:pill} displays the predicted dilepton 
mass spectrum for $D^+ \to \pi^+ \ell^+\ell^-$.  
Several distinct kinds of contributions are included.  
The short-distance SM component corresponds to the dashed 
line, which is seen to lie beneath the other two curves.
For reference, we cite the {\it inclusive} `short distance' 
branching ratio,
\be
{\cal B}r_{D^+\to X_u^+ e^+e^-}^{\rm (sd)} \simeq 2\times10^{-8} \ \ .
\label{sdbr}
\ee

\subsection{The long-distance component}\label{subsec:ld} 
\begin{figure}
\vskip .1cm
\hskip 3.0cm
\psfig{figure=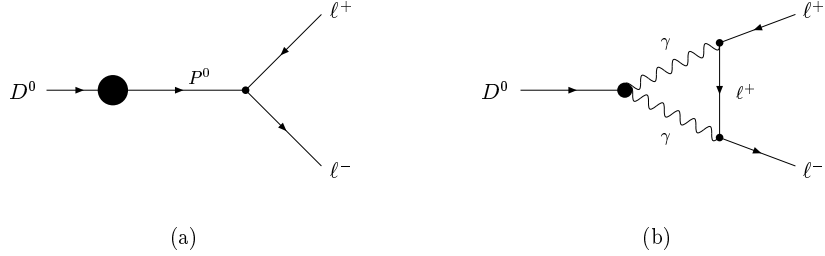,height=1.3in}
\caption{Long distance contributions to $D^0 \to \ell^+\ell^-$.\hfill 
\label{fig:ll}}
\end{figure}

\begin{figure}
\vskip .1cm
\hskip 2.0cm
\psfig{figure=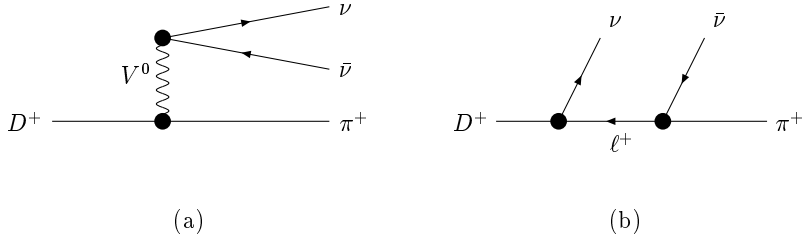,height=1.2in}
\caption{Long distance contributions to $D^+ \to 
\pi^+\ell^+\ell^-$.\hfill 
\label{fig:pll}}
\end{figure}

The long-distance component to a transition amplitude is 
often cast in terms of hadronic entities rather than 
the underlying quark and gluonic degrees of freedom.  
For charm decays, the long-distance amplitudes are typically 
important but difficult to determine with any rigor. There 
are generally several long-distance mechanisms for a given 
transition, {\it e.g.} as indicated for $D^0 \to \ell^+ \ell^-$ 
in Fig.~\ref{fig:ll} and for $D^+ \to \pi^+ {\bar\nu} \nu$ 
in Fig.~\ref{fig:pll}.  

Let us return to the case of $D^+ \to \pi^+ \ell^+\ell^-$ 
depicted in Fig.~\ref{fig:pill}.  The solid curve 
represents the {\it total} SM signal, summed over both SD and LD 
contribution.  In this case the LD component dominates, and 
from studying the dilepton mass distribution we can see what 
is happening.  The peaks in the solid curve must correspond 
to intermediate resonances ($\phi$, {\it etc}).  The corresponding 
Feynman graph would be analogous to that in Fig.~\ref{fig:pll}(a) in 
which the final state neutrino pair is replaced by a charged 
lepton pair.  One finds numerically that 
\be
{\cal B}r_{D^+\to \pi^+ e^+e^-}^{\rm (SM)} \simeq 
{\cal B}r_{D^+\to \pi^+ e^+e^-}^{\rm (\ell d)} \simeq 
2\times10^{-6} \ \ .
\label{ldbr}
\ee

\subsection{The Standard Model Predictions}\label{subsec:SM} 
Basing our analysis in part on existing literature,~\cite{lit} 
we have calculated both SD and LD amplitudes for a number 
of FCNC $D$ transitions.~\cite{bghp2}  Results are collected 
in Table~\ref{tab:br}.  As stated earlier, the current database 
for processes appearing in Table~\ref{tab:br} consists entirely 
of upper bounds (or in the case of $D^0 \to \gamma \gamma$ no data 
entry at all).  In all cases existing experimental bounds lie below 
the SM predictions, so there is no conflict between the two.  
For some ({\it e.g.} $D \to \pi \ell^+ \ell^-$) the gap between 
SM theory and experiment is not so large and there is hope for 
detection in the near future.  In others ({\it e.g.} $D^0 \to 
\ell^+ \ell^-$) the gap is enormous, leaving ample opportunity 
for signals from New Physics to appear.  This point is sometimes 
not fully appreciated and thus warrants some emphasis.  It is why, 
for example, attempts to detect $\Delta M_{\rm D}$ via 
$D^0$-${\bar D}^0$ mixing experiments are so important.

\begin{table}[t]
\caption{\small Standard Model predictions and current experimental 
limits for the branching fractions due to 
short and long distance contributions for various rare $D$ meson
decays.\label{tab:br}} 
\vspace{0.4cm}
\begin{center}
\begin{tabular}{|l|c|c|c|} \hline
Decay Mode & Experimental Limit & ${\cal B}r_{S.D.}$ & ${\cal B}r_{L.D.}$ 
\\ \hline
$D^+\to X_u^+ e^+e^-$ & & $2\times 10^{-8}$ & \\
$D^+\to\pi^+e^+e^-$ & $<4.5\times 10^{-5}$ & &$2\times10^{-6}$ \\
$D^+\to\pi^+\mu^+\mu^-$ & $<1.5\times 10^{-5}$ & &$1.9\times10^{-6}$ \\
$D^+\to\rho^+e^+e^- $ & $<1.0\times 10^{-4}$ & &$4.5\times10^{-6}$ \\
$D^0\to X_u^0+e^+e^-$ & & $0.8\times 10^{-8}$ & \\
$D^0\to\pi^0e^+e^-$ & $<6.6\times 10^{-5}$ &  &
$0.8\times10^{-6}$ \\
$D^0\to\rho^0e^+e^-$ & $<5.8\times 10^{-4}$ & & $1.8\times10^{-6} $ \\
$D^0\to\rho^0\mu^+\mu^-$ & $<2.3\times 10^{-4}$ & & $1.8\times10^{-6} $ \\
\hline
$D^+\to X_u^+\nu\bar\nu$ & & $1.2\times 10^{-15}$ & \\
$D^+\to\pi^+\nu\bar\nu$ & &  & $5\times 10^{-16}$ \\
$D^0\to\bar K^0\nu\bar\nu$ &  & & $2.4\times10^{-16}$ \\
$D_s\to \pi^+\nu\bar\nu$ & & & $8\times10^{-15}$ \\ \hline
$D^0\to\gamma\gamma$ & & $4\times10^{-10}$ & few~$\times 10^{-8}$ \\ \hline
$D^0\to\mu^+\mu^-$ & $<3.3\times 10^{-6}$ & $1.3\times 10^{-19}$ &
${\rm few}~\times 10^{-13}$ \\
$D^0\to e^+e^-$ & $<1.3\times 10^{-5}$ & $(2.3-4.7)\times 10^{-24}$ & \\
$D^0\to\mu^\pm e^\mp$ & $<8.1\times 10^{-6}$ & $0$ & $0$ \\ 
$D^+\to\pi^+\mu^\pm e^\mp$ & $<3.4\times 10^{-5}$ & $0$ & $0$ \\ 
$D^0\to\rho^0\mu^\pm e^\mp$ & $<4.9\times 10^{-5}$ & $0$ & $0$ \\ \hline
\end{tabular}
\end{center}
\end{table}

\section{New Physics Analysis}
At this time, there is a wide collection of possible 
New Physics models leading to FCNC $D$ transitions.  Among 
those considered in Ref.~1 are 
(i) Supersymmetry (SUSY): R-parity conserving, R-parity violating, 
(ii) Extra Degrees of Freedom: Higgs bosons, Gauge bosons, Fermions, 
Spatial dimensions, (iii) Strong Dynamics: 
Extended technicolor, Top-condensation.

Due to limitations of time (for the talk) and space (for this 
summary) we restrict most of our attention to the case of 
supersymmetry.  However, at the end we also make a few remarks 
on the topic of large extra dimensions.
The SUSY discussion divides naturally according 
to how the R-parity $R_{\rm P} $ is treated, where 
\begin{equation}
R_{\rm P} \ = \ (-)^{3(B-L)+2S} \ = \ \left\{
\begin{array}{cc}
+1 & {\rm (particle)} \\
-1 & {\rm (sparticle)} \ \ .
\end{array}
\right.
\end{equation}

\subsection{R-parity conserving SUSY}
R-parity conserving SUSY will contribute to charm FCNC amplitudes 
via loops.  For a penguin-like amplitude whose 
external lines are SM particles, the internal lines can be 
(i) gluino-squark pairs, (ii) charged Higgs-quark pairs or (iii) 
chargino/neutralino-squark pairs.  Case (i) is the one considered 
here as case (ii) will be suppressed for the same CKM reason 
as in the SM and case (iii) is relatively suppressed to (i) 
because the vertices are weak-interaction rather than
strong-interaction.  

To calculate R-parity conserving SUSY contributions, we 
employ the so-called {\it mass insertion approximation},~\cite{lhall} 
which is oriented towards phenomenological studies and is also 
model independent.  Let us first describe what is actually done 
and then provide a brief explanation of the underlying rationale.

\begin{figure}
\vskip .1cm
\hskip 3.5cm
\rotate[r]{\psfig{figure=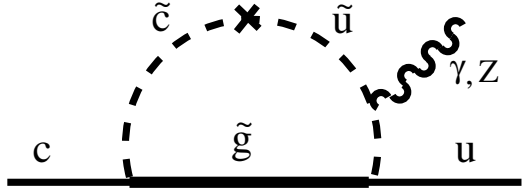,height=2.7in}}
\caption{A typical contribution to $c\to u$ FCNC transitions in 
the MSSM. The cross denotes one  mass insertion 
$(\delta^u_{12})_{\lambda\lambda'}$ and $\lambda,\lambda'$ 
are helicity labels.\hfill 
\label{fig:mssm}}
\end{figure}

In this approach, a squark propagator becomes 
modified by a mass insertion ({\it e.g.} the `$\times$' in 
Fig.~\ref{fig:mssm}) that changes the squark flavor.~\cite{lhall,susyfcnc} 
For convenience, one expands the squark propagator in powers of the 
dimensionless quantity $(\delta^u_{ij})_{\lambda\lambda'}$, 
\begin{equation}
(\delta^u_{ij})_{\lambda\lambda'}={(M^u_{ij})^2_{\lambda\lambda'}
\over M^2_{\tilde q}}\ \ , 
\label{delta}
\end{equation}
where $i\neq j$ are generation indices, $\lambda,\lambda'$ denote the
chirality, $(M^u_{ij})^2$ are the off-diagonal elements of the 
up-type squark mass matrix and $M_{\tilde q}$ represents the average 
squark mass.  The exchange of squarks in loops thus leads to FCNC
through diagrams such as the one in Fig.~\ref{fig:mssm}.  The role 
of experiment is to either detect the predicted (SUSY-induced) 
FCNC signal or to constrain the contributing 
$(\delta^u_{ij})_{\lambda\lambda'}$.

This topic is actually part of the super-CKM problem.  
If one works in a basis which diagonalizes the fermion mass 
matrices, then sfermion mass matrices (and thus sfermion 
propagators) will generally be nondiagonal.  As a result, 
flavor changing processes can occur.  One can use 
phenomenology to restrict these FCNC phenomena.  The 
$Q=-1/3$ sector has yielded fairly strong constraints but 
thus far only $D^0$-${\bar D}^0$ mixing has been used to limit 
the $Q=+2/3$ sector.  In our analysis, we have taken charm FCNCs 
to be as large as allowed by the $D$-mixing upper bounds.  

For the decays $D \to X_u \ell^+ \ell^-$ discussed
earlier in Sect.~\ref{subsec:sd}, the gluino contributions will 
occur additively relative to those from the SM and so we can 
write for the Wilson coefficients, 
\be
C_i = C_i^{\rm (SM)} + C_i^{\rm \tilde g} \ \ .
\label{wilson}
\ee
To get some feeling for dependence on the 
$(\delta^u_{12})_{\lambda\lambda'}$ parameters, we display 
the examples 
\be
C_7^{\rm \tilde g} \ \propto \ (\delta^u_{12})_{\rm LL}\ {\rm and} \ 
(\delta^u_{12})_{\rm LR}\ , 
\qquad 
C_9^{\rm \tilde g} \ \propto \ (\delta^u_{12})_{\rm LL}\ \ , 
\label{right}
\ee
whereas for quark helicities opposite~\footnote{We use the 
notation ${\hat C}$ for the associated Wilson coefficients.} 
to those in the operators of Eq.~(\ref{examples}), one finds 
\be
{\hat C}_7^{\rm \tilde g} \ \propto \ (\delta^u_{12})_{\rm RR}\ 
{\rm and} \ (\delta^u_{12})_{\rm LR}\ , 
\qquad 
{\hat C}_9^{\rm \tilde g} \ \propto \ (\delta^u_{12})_{\rm RR}\ \ .
\label{wrong} 
\ee
Moreover, the 
term in ${\hat C}_7^{\rm \tilde g}$ which contains 
$(\delta^u_{12})_{\rm LR}$ experiences the enhancement factor 
$M_{\tilde g}/m_c$. 

\begin{figure}
\vskip .1cm
\hskip 3.5cm
\rotate[l]{\psfig{figure=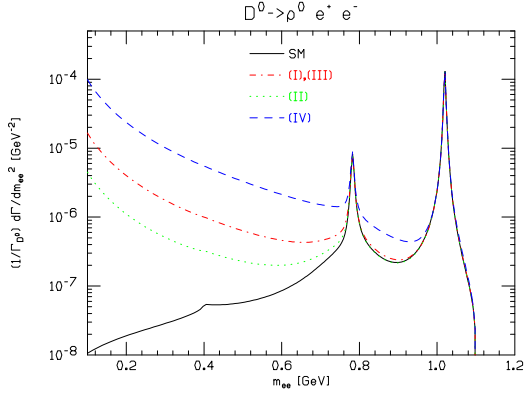,height=2.7in}}
\caption{Dilepton mass distributions for $D^0 \to \rho^0 e^+ e^-$ 
in the mass insertion approximation of MSSM.  The SM prediction 
(solid curve) is provided for reference and the MSSM curves 
refer to (i) $M_{\tilde g} = M_{\tilde q} = 250~{\rm GeV}$, 
(ii) $M_{\tilde g} = 2 M_{\tilde q} = 500~{\rm GeV}$, 
(iii) $M_{\tilde g} = M_{\tilde q} = 1000~{\rm GeV}$ and 
(iv) $M_{\tilde g} = M_{\tilde q}/2 = 250~{\rm GeV}$.\hfill
\label{fig:rhollmssm}}
\end{figure}

We have numerically studied the effects in $c\to u\ell^+\ell^-$ 
for the range of masses: 
(I) $M_{\tilde g}=M_{\tilde q}=250$~GeV, 
(II) $M_{\tilde g}=2\,M_{\tilde q}=500$~GeV,  
(III) $M_{\tilde g}= M_{\tilde q}=1000$~GeV and 
(IV) $M_{\tilde g}=(1/2)\,M_{\tilde q}=250$~GeV.  
For some modes in $D \to X_u \ell^+ \ell^-$, the effect of the 
squark-gluino contributions can be large relative to the SM 
component, both in the total branching ratio and for certain 
kinematic regions of the dilepton mass.  The mode 
$D^0 \to \rho^0 e^+ e^-$ is given in Fig.~\ref{fig:rhollmssm}.
This figure should make clear the importance of measuring 
the low $m_{\ell^+\ell^-}$ part of the dilepton mass spectrum.

\subsection{R-parity violating SUSY}
The effect of assuming that $R$-parity can be violated is to 
allow additional interactions between particles 
and sparticles.  Ignoring bilinear terms which are 
not relevant to our discussion of FCNC effects, we introduce 
the $R$-parity violating (RPV) super-potential of trilinear couplings, 
\be
{\cal W}_{{\not R}_p} =\epsilon_{ab}\left[
\frac{1}{2}\lambda_{ijk}L^a_iL^b_j\bar{E}_k
+\lambda'_{ijk}L_i^aQ^b_j\bar{D}_k 
+\frac{1}{2}\epsilon_{\alpha\beta\gamma}\lambda^{''}_{ijk}\bar{U}^\alpha_i
\bar{D}^\beta_j\bar{D}^\gamma_k \right]\ \ ,
\label{rpv1}
\ee
where $L$, $Q$, $\bar E$, $\bar U$ and $\bar D$ are the standard 
chiral super-fields of the MSSM and $i,j,k$ are generation indices.
The quantities $\lambda_{ijk}$, $\lambda_{ijk}'$ and 
$\lambda^{''}_{ijk}$ are {\it a priori} arbitrary couplings 
which total $9+27+9=45$ unknown parameters in the theory.  

For our purposes, the presence of RPV means that 
{\it tree-level} amplitudes become possible in which a virtual 
sparticle propagates from one of the trilinear vertices in 
Eq.~(\ref{rpv1}) to another In order to avoid significant FCNC signals 
(which would be in contradiction with current experimental 
limits), bounds must be placed on the (unknown) coupling parameters. 
As experimental probes become more sensitive, the bounds become 
ever tighter. In particular, the FCNC sector probed by charm decays 
involves the $\{ \lambda_{ijk}'\}$. Introducing matrices 
${\cal U}_{\rm L}$, ${\cal D}_{\rm R}$ to rotate left-handed up-quark 
fields and right-handed down-quark fields to the mass basis, we 
obtain for the relevant part of the superpotential 
\be
{\cal W}_{\lambda'} = {\tilde \lambda}_{ijk}' \left[ 
-\tilde{e}_L^i\bar{d}_R^k u_L^j - \tilde{u}_L^j\bar{d}_R^ke_L^i 
-(\tilde{d}_R^k)^*(\bar{e_L^i})^cu_L^j + \dots \right] \ \ ,
\label{rpv2}
\ee
where neutrino interactions are not shown and we define 
\be
\tilde{\lambda'}_{ijk}\equiv \lambda'_{irs} {\cal U}^L_{rj} 
{\cal D}^{*R}_{sk} \ \ .
\label{rpv3}
\ee
Some bounds on the $\{ \tilde{\lambda'}_{ijk} \}$ are already 
available from data on such diverse sources as 
charged-current universality, the ratio 
$\Gamma_{\pi\to e\nu_e}/\Gamma_{\pi\to \mu \nu_\mu}$, 
the semileptonic decay $D\to K \ell \nu_\ell$ , {\it etc}.~\cite{addr}
We have considered additional experimental implications of the preceding 
formalism:

(i) For the decay $D^+\to \pi^+ e^+ e^-$, we display 
the effect of RPV as the dot-dash line in Fig.~\ref{fig:pill}.  
Here, the effect is proportional to 
${\tilde\lambda}'_{11k}\cdot {\tilde\lambda}'_{12k}$ and 
we have employed existing limits on these couplings.
Although the effect on the branching ratio is not 
large, but the dilepton spectrum away from 
resonance poles is seen to be sensitive to the RPV contributions.  
This case is not optimal because the current experimental limit 
on ${\cal B}r_{D^+\to \pi^+ e^+e^-}$ is well above the dot-dash curve.

(ii) For $D^+\to \pi^+ \mu^+ \mu^-$, the current experimental 
limit on ${\cal B}r_{D^+\to \pi^+ \mu^+ \mu^-}$ actually 
provides the new bound 
\be
{\tilde\lambda}'_{11k}\cdot {\tilde\lambda}'_{12k} \le 0.004 \ \ .
\label{rpv4}
\ee

(iii) Another interesting mode is $D^0 \to \mu^+\mu^-$.
Upon using the bound of Eq.~(\ref{rpv4}) we obtain 
\begin{equation}
{\cal B}r^{\not R_p}_{D^0\to\mu^+\mu^-} < 3.5\times 10^{-6}\;
\left(\frac{\tilde{\lambda}'_{12k}}{0.04}\right)^2
\,\left(\frac{\tilde{\lambda}'_{11k}}
{0.02}\right)^2 \ \ .
\label{d2mu_bound} 
\end{equation}
A modest improvement in the existing limit on 
${\cal B}r_{D^0\to\mu^+\mu^-}$ will yield a new bound on 
the product ${\tilde\lambda}'_{11k}\cdot {\tilde\lambda}'_{12k}$. 

(iv) Lepton flavor violating processes are allowed by the 
RPV lagrangian.  One example is the mode $D^0 \to e^+ \mu^-$, 
for which existing parameter bounds predict 
\begin{equation}
{\cal B}r^{\not R_p}_{D^0\to\mu^+e^-} <0.5 \times10^{-6}\times
\left[ \left(\frac{\tilde{\lambda}'_{11k}}{0.02}\right)
\left(\frac{\tilde{\lambda}'_{22k}}{0.21}\right)
 + 
\left(\frac{\tilde{\lambda}'_{21k}}{0.06}\right)
\left(\frac{\tilde{\lambda}'_{12k}}{0.04}\right)
\right] \ \ .
\label{d2mue_2body}
\end{equation}
An order-of-magnitude improvement in 
${\cal B}r^{\not R_p}_{D^0\to\mu^+e^-}$ will provide a new 
bound on the above combination of RPV couplings.  

\subsection{Large Extra Dimensions}
For several years, the study of large extra dimensions 
(`large' means much greater than the Planck scale) has 
been an area of intense study.  This approach might hold the 
solution of the hierarchy problem while having verifiable 
consequences at the TeV scale or less.  Regarding the subject 
of rare charm decays, one's reaction might be to ask 
{\it How could extra 
dimensions possibly affect the decays of ordinary hadrons?}.
We provide a few examples in the following.

Suppose the spacetime of our world amounts to a $3+1$ brane 
which together with a manifold of additional dimensions 
(the bulk) is part of some higher-dimensional space.  
A field $\Theta$ which can propagate in a large extra 
dimension will exhibit a Kaluza-Klein (KK) tower of 
states $\{ \Theta_n \}$, detection of which would signal existence 
of the extra dimension.  Given our ignorance regarding properties 
of the bulk or of which fields are allowed to propagate 
in it, one naturally considers a variety of different models.

Assume, for example, the existence of an extra dimension of 
scale $1/R \sim 10^{-4}$~eV such that the gravitational field 
(denote it simply as $G$) alone can propagate in the extra 
dimension.~\cite{add}  There are then bulk-graviton KK states 
$\{ G_n\}$ which couple to matter.  In principle there will be 
the FCNC transitions $c \to u ~G_n$ and since the $\{ G_n\}$ 
remain undetected, there will be apparent missing energy.  
However this mechanism leads to too small a rate to be observable.

Another possibility which has been studied is that the scale 
of the extra dimension is $1/R \sim 1$~TeV and that SM gauge 
fields propagate in the bulk.~\cite{ant}  However, precision
electroweak data constrain the mass of the first gauge KK excitation to
be in excess of 4 TeV~\cite{tgr}, and hence their contributions to rare
decays are small~\cite{deshxd}.

More elaborate constructions, such as allowing fermion fields to 
propagate in the five-dimensional bulk of the Randall-Sundrum 
localized-gravity model~\cite{rs}, are currently being actively 
explored.~\cite{frank}  Interesting issues remain and a good 
deal more study deserves to be done.

\section{Concluding Remarks} 
In the Standard Model, FCNC processes are suppressed.  
Both photon and Z-boson SM vertices are flavor diagonal, 
so tree level diagrams involving virtual propagation of 
these particles will not contribute to FCNC processes.  
One must instead consider loops.  As regards FCNC processes, 
SM loops are largest for kaon and B-meson transitions 
due mainly to the large $t$-quark mass and also to 
favorable CKM dependence.  The upshot is 
that studies of FCNC processes for charm have lagged behind 
those of the other flavors.  There is of course a basic irony 
in this -- it is precisely because SM signals are expected to be 
so small for charm FCNC that the opportunity for evidence of 
New Physics to emerge becomes enhanced relative to the other 
flavors.  The situation begs for experimental input.

\section*{Acknowledgments}
The work described in this talk was supported in part by the National 
Science Foundation under Grant PHY-9801875.

\end{document}